\documentclass[preprint,groupedaddress,floatfix,%
nofootinbib]{revtex4}

\usepackage{euscript}
\usepackage{dcolumn}
\usepackage{graphicx}
\usepackage{epsfig}
\usepackage{hyperref}
\usepackage{graphicx}
\usepackage{dcolumn}
\usepackage{longtable}
\usepackage{times}
\usepackage{amsmath,amssymb,bm}
\usepackage[english]{babel}
\usepackage[latin1]{inputenc}
\usepackage{times}
\usepackage[T1]{fontenc}
\usepackage{color}
\usepackage{fancybox}
\usepackage{sidecap}
\usepackage{graphicx}
\usepackage{epsfig}
\usepackage{psfrag}
\usepackage{bbm}
\usepackage[english]{babel}
\usepackage[latin1]{inputenc}
\usepackage{times}
\usepackage[T1]{fontenc}
\usepackage{color}
\usepackage{fancybox}
\usepackage{sidecap}
\usepackage{psfrag}
\usepackage{bbm}
\usepackage{wasysym}
\usepackage{slashed}
\usepackage{tikz}
\usepackage{euscript}

\newcommand{\mathd}{\mathrm{d}}
\newcommand{\mathi}{\mathrm{i}}
\newcommand{\mathe}{\mathrm{e}}

\begin{document}

\title{The Cosmological Constant: A Categorical View}
\author{Fen Zuo\footnote{Email: \textsf{18802711412@139.com}}}
\affiliation{~}

\begin{abstract}

Some theoretic results related to the cosmological constant, obtained in the 80's and 90's of last century,
are reviewed. These results exhibit some interesting underlying pattern when viewed from a category-theoretic perspective.
In doing this, we illustrate how Baez and Dolan's Periodic Table of $k$-tuply monoidal $n$-categories can serve as a basic framework of a quantum spacetime structure. Explicitly, we show how Einstein gravity (with the cosmological constant turned on) emerges when we de-categorify certain $2$-tuply monoidal $2$-categories properly, as Crane proposed a long time ago. In particular, we find that some recent work on Vertex Operator Algebras and 4-manifolds fits nicely into this framework.

\end{abstract}
 \maketitle

\tableofcontents

\section{Introduction}

Perhaps Louis Crane is the first person to recognize the necessity of category theory in constructing a quantum theory of gravity~\cite{Crane:1993vs}. At a time when higher-category theory was still in its infancy stage, he realized that categorification~\cite{Crane:1994ty} is an inevitable step in quantum gravity~\cite{Crane:1995qj}. Unfortunately, as higher category theory is slowly developed, its importance in quantum gravity was never fully appreciated by the physics community. Nevertheless, there is at least one person, John Baez, who has kept and further developed Crane's ideas. Baez has proposed to use higher category theory as a language to construct a quantum spacetime for a long time, starting from the pioneer paper~\cite{Baez:1995xq}, in the later series of papers~\cite{Baez:1999in}\cite{Baez:2004pa}, and especially in the talk~\cite{Baez:2006A}. Here I want to follow Crane and Baez's approach a little further, to see how the cosmological constant could be accommodated in such a framework. I will assume that the readers are somehow familiar with the basics of $n$-categories, or at least, have read the pioneer papers~\cite{Baez:1995xq}\cite{Baez:1998} and the further interpretation~\cite{Baez:2006B}. For physicists, Baez's nice review paper~\cite{Baez:2009}, and many early pedagogical articles on his blog, ``This Week's Finds in Mathematical Physics''~\cite{TWF}, are much easier to follow.

I will first review the historical progresses made on the problem and show how to reinterpret them from the categorical point of view. Then I will show that, some recent progresses involving Vertex Operator Algebras~(VOAs) and 4-manifolds~\cite{Gadde:2013sca}\cite{Gukov:2017}\cite{Gukov:2018}, originating from the
string/M-theory framework, could be incorporated into such a program. In particular, I want to point out that these progresses could help reveal the categorical origin of the realistic cosmological constant. It is gratifying that the importance of categorification~\cite{Crane:1994ty} in studying 4-manifolds was recognized recently in~\cite{Gukov:2017}. However, its physical importance in constructing quantum gravity~\cite{Crane:1995qj} is still largely ignored. I hope that this paper could serve as a remedy for this.

\subsection{The notation}

The cosmological constant in an arbitrary $n$ dimensional spacetime is commonly introduced through an additional constant term to the Hilbert-Einstein action:
\begin{equation}
S_{\Lambda \text {GR}} \sim \frac{1}{2\kappa_n} \int \mathd ^n x \sqrt{\sigma g} (R-2 \Lambda_n).
\end{equation}
with Einstein's constant $\kappa_n=8\pi G_n/c^2$, and $\sigma=+1(-1)$ for Euclidean (Lorentzian) signature. Note the sign convention in front of the constant term.
It is convenient to rewrite the cosmological constant as
\begin{equation}
\Lambda_n=\frac{(n-1)(n-2)}{2L_n^2}.\label{eq:L}
\end{equation}
We restrict $\Lambda_n$ to be real, so $L_n$ is real for $\Lambda_n>0$ and pure imaginal for $\Lambda_n<0$.
We also define the corresponding Planck length as
\begin{equation}
l_n^{n-2}=\frac{8\pi \hbar G_n}{c^3}.
\end{equation}
One may doubt how the Planck constant $\hbar$ is introduced in pure gravity. A physical way to do this is through the Ponzano-Regge model in 3D~\cite{Ponzano-Regge}, or the Rovelli-Smolin quantization of geometric quantities in 4D, see~\cite{Rovelli:1994ge} for the original formulation. In this kind of kinematical quantization procedure, it is always the Planck length $l_n$, rather than the gravitational constant $G_n$ alone, that naturally appears in the quantized theory. Anyhow, it is introduced artificially, just as in ordinary quantum mechanics. We will discuss this in detail later.

\section{$3$D}
We start with the 3D gravity, which may be relatively more familiar to the readers, due to the pioneer work
of Witten~\cite{Witten:1988}. We treat the Euclidean and the Lorentzian case separately, since the underlying mathematical structure can be quite different, actually.

\subsection{Euclidean case}

Although in~\cite{Witten:1988} the author mainly deals with the Lorentzian case~(when the cosmological constant is turned on), the procedure can be easily generalized to the Euclidean case~(see, e.g.,~\cite{Regge:2000wu}\cite{Carlip:1997}\cite{Carlip:2014bfa}.
When a cosmological constant $\Lambda_3$ is introduced, the Euclidean gravity can be expressed as
a ``doubled"/non-chiral Chern-Simons-Witten~(CSW) theory~\cite{Witten:1988hf}, with the gauge group $\text{SU}(2)\times \text{SU}(2)$. The gauge fields are constructed as the combination of spin connection and the frame field:
\begin{equation}
A_{\pm}=\omega \pm \sqrt{\Lambda_3} e.
\end{equation}
Notice that for these fields to be real, $\Lambda_3$ has to be positive. Here I must mention that, the geometric meaning of these combinations are unclear to me. The 3D Einstein-Hilbert action, with the cosmological term, now becomes
\begin{equation}
S_{\Lambda \text {GR}}= S_{\rm CS}(A_+) - S_{\rm CS} (A_-),
\end{equation}
with the definition of the CSW action~\cite{Witten:1988hf}:
\begin{equation}
S_{\rm CS}(A)=\frac{k}{4\pi}\int_{M_3} {\rm Tr}\left(A\wedge \mathd A+\frac{2}{3}A\wedge A\wedge A\right).
\end{equation}
The coupling $k$ is related to the cosmological constant as~\cite{Witten:1988}:
\begin{equation}
k=\frac{1}{4G_3\sqrt{\Lambda_3}}=\frac{2\pi L_3}{l_3}. \label{eq:k3E}
\end{equation}
From this expression, it is clear that $k$ measures the cosmological length (perimeter) in the Planck unit.

One may doubt the above ``physical'' procedure, which heavily depends on the action functional.
A ``mathematical'' version, parallel to the above procedure, can actually be established in a rigorous way. Here we also sketch such a derivation. I believe this is very well reviewed in~\cite{Regge:2000wu}. In 1961 Regge invented a discrete formulation of general relativity, which is now
called Regge calculus~\cite{Regge:1961}. In this approach, the smooth manifold is replaced by the corresponding triangulation, in which the codimension-2 simplices, called hinges or bones, play an important role. The measure on the hinges plays the role of metric, while the deficit angles localized on them play the role of the curvature. Still, these hinge measures and deficit angles are classical variables. Especially, in 3D the hinge measures are simply the lengths of the edges of the triangulation. Much like the quantization of angular momentum in quantum mechanics, we could quantize these edge lengths. In this way the Ponzano-Regge model is constructed in 1968~\cite{Ponzano-Regge}, which is a discrete path integral, or
state sum, involving the elements of the representation category of SU$(2)$. In particular, the $6j$-symbols appear explicitly in the state sum. Based on a conjectured, and proved much later~\cite{Roberts:1995}, asymptotic formula for the $6j$-symbols, the Ponzano-Regge model is shown to reproduce the Regge action~\cite{Regge:1961} in the classical limit.
The structure used in Ponzano-Regge model is independently invented by Penrose a few years later, with the name ``spin network''~\cite{Penrose:1971}. The problems with the Ponzano-Regge model, and Penrose's spin network, is that the state sum could be divergent, since the spins can take arbitrary large values. Until almost three decades later,
a regularized version of the Ponzano-Regge model is rigorously constructed by Turaev and Viro~\cite{Turaev:1992hq}, based on the representation category of the quantum group $U_{q}\mathfrak{s}\mathfrak{l}_2$. In particular, the ``quantum'' $6j$-symbols are involved in the improved state sum. In the limit when the deformation parameter
$q\to 1$, the Turaev-Viro theory reproduces the Ponzano-Regge model.

Then the natural question is, how is the Turaev-Viro theory related to the CSW formulation of 3D gravity? While 3D Euclidean
gravity could be formulated by a ``doubled'' CSW theory with gauge group $\text{SU}(2)\times \text{SU}(2)$, the Turaev-Viro theory is also a ``doubled'' version of a previously constructed topological quantum field theory~(TQFT), Reshetikhin-Turaev theory~\cite{Reshetikhin:1991}. It turns out that the Reshetikhin-Turaev theory is equivalent to the geometric quantization of CSW theory, see the discussion in~\cite{mathoverflow:86792} and \cite{mathoverflow:178113}, and references therein. One of the key steps in establishing this equivalence is the equivalence of the representation category on both sides. Namely, in the Reshetikhin-Turaev approach, we have the representation category of the quantum group $U_{q}\mathfrak{sl}_2$. While in the geometric quantization of CSW theory, we have the modules of the
affine Lie algebra $\widehat{\mathfrak{s}\mathfrak{u}(2)}_k$. It turns out that they are equivalent as
modular tensor categories~(MTCs)~\cite{Kazhdan-Lusztig:I,Finkelberg:1996,Finkelberg:2013}:
\begin{equation}
\text{Rep}(U_{q}\mathfrak{s}\mathfrak{l}_2)\cong \EuScript{C}_{\widehat{\mathfrak{s}\mathfrak{u}(2)}_k},\label{eq:3-2}
\end{equation}
with
\begin{equation}
q=\mathe^{\mathi \pi/(k+2)}.
\end{equation}
Together with the relation (\ref{eq:k3E}), this gives the relation between $q$ and $\Lambda_3$. When $q\to 1$, the cosmological constant vanishes. Thus the Turaev-Viro theory introduces a nonzero cosmological constant into the Ponzano-Regge model, which regularizes the infinities appearing in the latter. The relation between the ``$q$'' parameter in the quantum group and the cosmological constant is discussed in~\cite{TWF:183}. In fact, I decided to write this note partly because I want
to make the discussion in the blog more thorough. In particular, I want to find out if the realistic cosmological constant, in the physical 4 dimensional spacetime, could be induced from a similar deformation quantization procedure. This will be the main content of our later sections.

So we see that even quantization of 3D Euclidean gravity is not simple. Here I repeated many of the details just to emphasize that, in either way, the quantization is a two-step procedure: either you introduce $\hbar$ to the theory first, and then deform it by the parameter $q$, as in the Ponzano-Regge and Turaev-Viro framework; or you introduce $k$ first, and then do the ordinary quantization with the Planck constant $\hbar$, as in the geometric quantization of CSW formulation. Due to the unexpected equivalence (\ref{eq:3-2}), the two quantization
procedures turn out to be equivalent. But why should such an equivalence of representation categories exist at all, as questioned in~\cite{mathoverflow:177519}? Recent work of Feigin and Gukov~\cite{Gukov:2018} gives a very interesting plausible explanation to this. Their work even suggests a unified quantization procedure, in which the cosmological constant and the Planck constant are treated more symmetrically. This will be discussed in the later sections, and now let us turn our focus to the Lorentzian 3D gravity.

\subsection{Lorentzian case}

In the Lorentzian case the quantum group approach seems to be unavailable. Fortunately, we could still employ the CSW approach, at least on the physical level of rigor. This can be easily obtained through a kind of ``Wick rotation''~\cite{Witten:1988}\cite{Regge:2000wu}\cite{Carlip:1997}\cite{Carlip:2014bfa} from the Euclidean one, and we are led to a doubled CSW theory with gauge group $\text {SO}(2,1)\times \text {SO}(2,1)$. The gauge fields are defined through
\begin{equation}
A_{\pm}=\omega \pm \sqrt{-\Lambda_3} e.
\end{equation}
Correspondingly, the coupling is now
\begin{equation}
k=\frac{1}{4G_3\sqrt{-\Lambda_3}}=\frac{2\pi \sqrt{-1}~L_3}{l_3}.\label{eq:k3L}
\end{equation}
To make sure that $A_{\pm}$ and $k$ are real, $\Lambda_3$ must be negative, which in turn results in a positive $k$. So we have an Anti-de Sitter~(AdS) spacetime in general. According to the definition (\ref{eq:L}), in this case $L_3$ is pure imaginary.
Upon geometric quantization, the relevant data would be the modules of the affine Lie algebra $\widehat{{\mathfrak s \mathfrak o}(2,1)}_k$.

In summary, three-dimensional gravity is always described by a (doubled/non-chiral) CSW theory with a positive coupling $k$,
 which I believe is the essential parameter here. Depending on the gauge group, the spacetime solution can be of Euclidean or Lorentzian signature. Once the signature is chosen, the sign of the cosmological constant is completely fixed. In the former case the corresponding solution can only be a sphere, and an $\text{AdS}_3$ spacetime in the latter.

\section{2D}

\subsection{Euclidean case}

Before I go on to our main topic in 4 dimensional spacetime, I believe some discussions on the properties of
possible 2D boundaries/domain walls in the above 3D quantum gravity are necessary. In a previous note~\cite{Zuo:2017hii} I have discussed this quite a lot, so here I only sketch the main results. The relation between the bulk CSW theory and the corresponding boundary conformal field theory~(CFT), namely a chiral part of the Wess-Zumino-Witten~(WZW) model, has been indicated in the original paper~\cite{Witten:1988hf}. However, a systematical
construction of the full theory, consistent on arbitrary surfaces, is far from clear. Until around 2005, the full rational conformal field theory~(RCFT) was finally rigorously constructed by Fuchs-Runkel-Schweigert~(FRS) (see~\cite{Fuchs:2001am} for a sketch of the procedure, and~\cite{Runkel:2005qw} for a review). The key idea in the FRS approach is the separation of the local complex-analytic and the global topological data. While the former is given
by proper VOA, the latter is described by the corresponding module structure. To avoid the technical complexities involved
in the FRS approach, we use some arguments from condensed matter physics to illustrate the main idea.

We know conformal theories are scale invariant, and thus gapless in the condensed matter language. But a special property of RCFTs is that, they are actually gappable~\cite{Kong:2018rig}. We already mention a chiral WZW model as the
boundary theory of a bulk CSW theory. When we have two opposite boundaries, the two boundary theories are of opposite chiralities. Then we compress the system to merely a single surface, obtaining a full WZW model. Now the surface theory is simply embedded in the trivial phase, and thus gappable~\cite{Kong:2018rig}. It is argued and conjectured in~\cite{Kong:2018rig} that all RCFTs can be obtained in this way, and thus are all gappable. Gappability reduces the construction of RCFTs to a pure topological
problem, which in 2D is classified around the same time~\cite{Fukuma:1993hy}\cite{Moore-Segal:2001,Segal:2001}\cite{Lauda:2006mn}.
Still there is a big difference between the two. In 2D TQFT the relevant monoidal category is the category of vector spaces, while in 2D RCFT one has the representations of rational VOAs, which has the structure of MTCs as proved in Huang's series of work~(for a review, see~\cite{Huang:2004dg}).
Then by analogy with the Fukuma-Hosono-Kawai~\cite{Fukuma:1993hy} and Lauda-Pfeiffer~\cite{Lauda:2006mn} construction, the FRS approach reduces to selecting a Fronenius algebra in the MTC of the VOA, together with its full center in the corresponding Drinfeld/monoidal center of the MTC. For our present case of $\text{SU}(2)$ WZW model, the Frobenius algebra and its full center can be explicitly written down. We label the irreducible representations $R_i$ of $\widehat{\mathfrak{su}(2)}_k$ by the spins $0\le i\le k/2$, and denote the conjugate representation as $R_i^\vee$. Then the Frobenius algebras are of the form
 \begin{equation}
 A=R_i \otimes  R_i^\vee, \label {eq:A}
 \end{equation}
considered as an object in the representation category $\EuScript{C}_{\widehat{\mathfrak{su}(2)}_k}$.
The full center is the Morita-equivalent class of them
\begin{equation}
Z(A)=\oplus_{0\le i\le k/2} R_i\times R_i^\vee, \label{eq:ZA}
\end{equation}
considered as an object in the Drinfeld center $\EuScript{C}_{\widehat{\mathfrak{su}(2)}_k}\boxtimes (\EuScript{C}_{\widehat{\mathfrak{s}\mathfrak{u}(2)}_k})^-$. This turns out to be the diagonal modular-invariant in the WZW model.

I have repeated the above discussion for two reasons: first, the Frobenius algebra and its full center in the FRS construction
has a pivotal role in turning the 4D topological theory into general relativity, at least in the Euclidean case~\cite{Zuo:2017hii}. This is the essence of the Barret-Crane model of 4D quantum gravity~\cite{Barrett:1997gw}. Secondly,
the separation of the local complex-analytic and global topological data in the FRS construction may persist in the Lorentzian case, in which the VOA is not rational any more. More importantly, this could even be an important property of the underlying quantum theory of gravity.

\subsection{Lorentzian case}
In the Lorentzian case we could still consider the corresponding VOA $\widehat{{\mathfrak s \mathfrak o}(2,1)}_k$, but the representation category is much more complicated. However, we hope a similar full center/diagonal modular invariant still exists.

Instead of going deep into the detailed representation theory, we would like to just point out that the corresponding $\text{SO}(2,1)$,
or $\text{SL}(2,\mathbb{R})$ WZW model is essential for the entropy of the black hole in the bulk. As shown by Carlip~\cite{Carlip:1994gy},
the black hole microstates are parameterized by the ``colored'' partition of a large excitation level~\cite{Carlip:1998qw,Carlip:2000nv}.
Therefore, the entropy could be easily derived using an extended Hardy-Ramanujan asymptotic formula.
The number of ``colors'' corresponds to the classical value of the central charge
\begin{equation}
c_{\rm{WZW}}=\frac{3k}{k-2}\to 3,\quad k\to \infty. \label{eq.c-aymp1}
\end{equation}
Interestingly, there is an alternative derivation based on the Virasoro VOA by Strominger~\cite{Strominger:1997eq}.
The derivation is in some sense indirect, and employs the Cardy asymptotic formula for a large Virasoro central charge.
Much later~\cite{Carlip:2005zn},
Carlip realized that the two are related through a gauging procedure leading WZW model to Liouville field theory.
Especially, the central charge of the latter is shown to be
\begin{equation}
c_{\rm{Liou}}\to 6k , \quad k\to \infty. \label{eq.c-aymp2}
\end{equation}
Based on the later series of works~\cite{Teschner:2005}\cite{Schomerus:2007}\cite{Giribet:2008}, we now know that the exact relation between the two theories includes the identification
\begin{equation}
k-2=b^{-2}.\label{eq:kb}
\end{equation}
And we know the exact central charge of the Liouville theory is
\begin{equation}
c_{\rm{Liou}}=1+6(b+b^{-1})^2.
\end{equation}
In the large $k$ limit, the limiting value of Strominger and Carlip (\ref{eq.c-aymp2}) is recovered. The relation between the
$\text{SO}(2,1)$ WZW model and the Liouville theory plays an important role in this paper and will be discussed in detail later.

\section{4D}

The situation in 4D is much more complicated and I believe is still not quite clear right now. The story started with the discovery of the
so-called Wheeler-de Witt~(WdW) equation~\cite{DeWitt:1967yk}\cite{Wheeler:1968}, which is an attempt to quantizing Einstein gravity in the canonical approach. The resulting
equation, as an analogue of Shr\"{o}dinger equation in quantum mechanics, is a complete mess in metrical formulation. The problem got simplified
20 years later, when the chiral connection~\cite{Ashtekar:1986yd}, instead of the metric, is used as the dynamical
variable. Expressed with these new variables, the WdW equation simply requires that they be flat~\cite{Jacobson:1988}.
The true meaning of the WdW equation was revealed almost ten years later by Smolin~\cite{Smolin:1995vq}, when a nonzero cosmological constant is turned on. Smolin shows that the WdW equation reveals the relation between a bulk BF theory~\cite{Horowitz:1989}\cite{Blau:1991} and the corresponding CSW theory at the boundary. This should not be quite surprising, since the 3D CSW action has a natural definition through the 2nd Chern class in 4D. Actually, this kind of relation has already been anticipated in~\cite{Crane:1995qj}, in which it was conjectured: ``The universe is in the CSW state''. Indeed, as further illustrated by Baez~\cite{Baez:1995ph}, the WdW equation with a cosmological constant
is solved by the CSW state~\cite{Kodama:1990}. A nice explanation of Smolin's discovery was given in~\cite{TWF:56,TWF:57}.

With Smolin's reinterpretation of the WdW equation, a natural question emerges: if the WdW equation is topological and encodes no dynamics of general relativity, how to implement or restore them? I think this question is much more complicated than it may look like. I will review the historical progresses made since then, and add some comments from the categorical viewpoint. Again I distinguish the Euclidean and Lorentzian cases.

\subsection{Euclidean case}

A direct way of implementing the dynamics of general relativity is to go down to lower dimensional geometrical objects, which was initiated already in Smolin's original paper~\cite{Smolin:1995vq}. It was concretely realized finally in the Barrett-Crane model~\cite{Barrett:1997gw}. In fact, the Barrett-Crane model is far more than simply a ``model'', which was explained in my recent note~\cite{Zuo:2017hii}. In the following I will sketch the main idea of the model, and also repeat some content in the note~\cite{Zuo:2017hii}.

\subsubsection{Barrett-Crane Model}

Once we have understood the true meaning of the WdW equation, it becomes inevitable to view general relativity as close cousins of BF theory. It has long been known that, the Pleba\'{n}ski Lagrangian~\cite{Plebanski:1977} for general relativity has the form of BF theory with an additional constraint. The constraint was analyzed in detail in~\cite{Barrett:1997gw}.
Viewing the $B$-field in the BF theory as a bivector made from the framing fields $e$, the constraint simply says that
$B$ should be a simple bivector of the form $e\wedge e$. Thus the constraint is named ``simplicity condition''. To specify the condition concretely, one should first define the BF theory in a rigorous way. Similar as the Reshetikhin-Turaev realization of CSW theory, one would like to formulate the BF theory in a combinatorial way. In~\cite{Baez:1995ph} it was argued that Crane and Yetter's topological state sum~\cite{Crane:1993if}, using the same data as the Reshetikhin-Turaev theory, provides such a formulation. Based on the Crane-Yetter state sum, Barrett and Crane then found a beautiful way to impose the simplicity condition, at least in the Euclidean case, as follows.

First we could identify the bivectors with the the elements in the algebra $\mathfrak{so}(4)$. The rotation group $\text{SO}(4)$, or its spin cover $\text{Spin}(4)$, has the decomposition
\begin{equation}
\text{Spin}(4) \cong \text{SU}(2)\times \text{SU}(2).\label{eq.Spin}
\end{equation}
The Crane-Yetter state sum could be constructed independently
with the representation data of the ``quantum deformed'' version of each subgroup. To recover Turaev-Viro formulation
of 3D gravity, the deformation parameters are chosen to be related by an inversion. Ignoring the deformation, we see that
the representations here could be arbitrary. In particular, since the chiral and anti-chiral
parts could be independently constructed, the representations for each subgroup are completely unrelated. Barrett and Crane
then shows, the simplicity condition combines the two parts closely together. This is done as follows. Corresponding to
the above decomposition, the bivectors are decomposed into the self-dual and anti-self-dual parts. In four dimension, a necessary and sufficient condition for a bivector to be simple is, it should be an equal-norm combination of its self-dual and anti-self-dual parts. This inspires Barrett and Crane to restrict the representations of $\text{Spin}(4)$ in the Crane-Yetter state sum to be of the diagonal form $(R_i,R_i^\vee)$. Since bivectors integrated over surfaces give the areas, such a choice was dubbed ``left-handed area = right-handed area''~\cite{Reisenberger:1998fk}.
As a punchline we may say, although Ashtekar's chiral connections simplify the WdW equation of general relativity miraculously, gravity is essentially non-chiral. Or in another way, essential variables of gravity separate into chiral
and anti-chiral parts at the level from 4D to 3D, but recombine non-chirally at the level from 3D to 2D.

\subsubsection{Anyon-Condensation Picture}

However, the beautiful structure hidden in the Barrett-Crane model is not fully appreciated. For example, in~\cite{Regge:2000wu} the authors claim that ``the Barrett-Crane amplitude is NOT independent of triangulation...''.
Even Barrett himself believes that in the model ``the constraints on objects and morphisms typically spoil the monoidal
product...''\cite{Barenz:2016nzn}. Actually all these statements are incorrect. These are clearly clarified
in my recent paper~\cite{Zuo:2017hii}. We may summarize the properties of the Barrett-Crane model by simply saying: it is full of proper
categorical virtues! To be more convincing, let me repeat some details in~\cite{Zuo:2017hii} as follows.

First we could use the equivalence (\ref{eq:3-2}) to change freely from the representations of the quantum group to the
corresponding affine Lie algebra. Then we will see all the ingredients in the Barrett-Crane model are exactly the topological data in the FRS formalism of RCFTs explained in the previous section. As I mentioned, in the rational case the conformal theories are actually gappable, and thus topological essentially. So we can switch to an equivalent picture in the field of topological phases, in which it is described through a process of ``anyon condensation''~\cite{Kong:2013aya}.
In particular, the anyon condensation process can be completely described in the categorical framework~\cite{Kong:2013aya}.
In this picture, the Barret-Crane model may be characterized through the following conjecture:\\

\textbf{Conjecture I:  Diagonal anyon-condensing the fully-extended (4-3-2-1-0), non-chiral Crane-Yetter theory
results in the Barret-Crane model, which is the combinatoric formulation of Euclidean general relativity (with the
cosmological constant).}\\

Explicitly, the procedure goes like this. First construct the fully-extended (4-3-2-1-0) Crane-Yetter theory from the data of
$\EuScript{C}_{\widehat{\mathfrak{su}(2)}_k}\boxtimes (\EuScript{C}_{\widehat{\mathfrak{su}(2)}_k})^-$. This theory is already classical at the top
level~\cite{Crane:1993cm}\cite{Roberts:1995}. This is in accordance with the corresponding analysis in the BF theory~\cite{Baez:1995ph}.
In the field of topological phases, it is supposed to describe the bulk of a topological insulator~\cite{Walker-Wang:2011}.
Notice that in~\cite{Walker-Wang:2011} an equivalent formulation of the Crane-Yetter theory is given in Hamiltonian framework.
At the next level, one may view it as a non-chiral CSW theory, or the Turaev-Viro theory. In either formulation
it contains nontrivial topological excitations and gives rise to nontrivial quantum topological invariants. In topological phases, this corresponds to the so-called (intrinsic) topological order ~\cite{Wen:1989iv}\cite{Wen:2004}. To condensing it completely, one has to choose a new vacuum of the condensed phase, such that all the topological excitations have non-trivial braiding with it. In the present case, this turns out to be the
full center/diagonal modular invariant (\ref{eq:ZA}), which we repeat here
\begin{equation}
Z(A)=\oplus_{0\le i\le k/2} R_i\times R_i^\vee, \label{eq:ZA2}
\end{equation}
This object fully captures the properties of the 3-cells, or tetrahedra, in the Barrett-Crane model. Maybe this needs a little explanation.
Remembering that in the Barrett-Crane model, we represent a tetrahedron in the dual picture by a four-edge vertex, and then split the vertex arbitrarily into
two three-edge vertices. The information of the interior of the tetrahedron is encoded on intermediate edge, which is colored by those diagonal representation $(R_i, R_i^\vee)$, with $0\le i\le k/2$. But as a special state sum, we have to sum up the contributions of all of them finally. This is equivalent to coloring the intermediate edge with the vacuum object $Z(A)$, since the external objects couple to the composite object through individual couplings with its sub-objects. I believe this is most obvious in the formulation of the Crane-Yetter state sum through
Skein theory~\cite{Roberts:1995}. In the Barrett-Crane model we should further specify the triangles. In the dual picture these are the external edges, colored with a specific $(R_i, R_i^\vee)$. But now no summation is needed if the corresponding triangle indeed appears on the 2D boundary. This corresponds to the vacuum object in the boundary phase, which is composed of the condensed anyons. Actually it is exactly the Frobenius algebra mentioned before (\ref{eq:A}):
 \begin{equation}
 A=R_i \otimes  R_i^\vee, \label{eq:A2}
 \end{equation}
The indices could be arbitrarily chosen in the allowed range $0\le i\le k/2$ , since all of them are Morita-equivalent.
Thus the Barrett-Crane model is fully captured by the anyon-condensation process in 3D, as we conjectured above.

Then one may ask, what are the advantages of the anyon-condensation picture of the Barrett-Crane model?
First, using the anyon-condensation framework we may easily disprove the criticisms made in~\cite{Regge:2000wu} and ~\cite{Barenz:2016nzn}. The criticism in the latter is easy to refute: the monoidal structure is fully preserved
in the condensation process, as illustrated clearly in~\cite{Kong:2013aya}. But the criticism in~\cite{Regge:2000wu} says that
``the Barrett-Crane amplitude is NOT independent of triangulation...'', is it true?

This is a very interesting question and can be analyzed from either the anyon-condensation framework or FRS formalism of RCFT.
We know that the fully-extended Crane-Yetter theory is completely triangulation-independent, at each level. The anyon condensation formalism actually preserves all the topological properties of the original theory, therefore should also be triangulation independent at the corresponding level. In particular, as in shown in~\cite{Kong:2013aya}, the condensed 3D bulk theory is a new topological phase described by the category of local $Z(A)$-modules, which is still a MTC. In the present case it is trivial and the 3D theory is a trivial topological phase. Therefore the relative 4D theory is also trivial and triangulation independent. The 2D boundary theory is described by the spherical fusion category of $A$-bimodules, and again topological and triangulation independent. One may dig down to even lower dimension,
which we will not try here. Now we turn to the FRS formalism, which could be viewed as the generalization of the traditional 2D lattice
topological theory. The lattice TQFT, in the open/closed extended version, is captured by the ordinary Frobenius algebra together with its center. The Frobenius property ensures the 2D triangulation independence. The same conclusion could be obtained by similar derivation, as shown in~\cite{Fuchs:2001am}. In essence, this is due to the gappability property of RCFT as mentioned above. One may also go down to lower dimension,
which will not be pursued here. In summary, in either the anyon-condensation or the FRS picture of the Barrett-Crane model, it is triangulation independent from 4D to 2D.

\subsubsection{Gravity From Condensation?}

After clarifying these doubts on the Barrett-Crane model, I now want to explain why it is conjectured to be a combinatoric formulation of Euclidean general gravity. Historically many ingredients of TQFT appear in the quantization process of gravity~(see~\cite{Barrett:1995} for a detailed
discussion of their common properties), yet people are
fully aware that there are huge difference between TQFT and general relativity. I believe this is most clearly clarified in~\cite{Baez:1999in}.
Therefore, even if the Barrett-Crane model is built out by imitating the classical deformation
from BF theory to general relativity, and we suppose that it indeed succeeds, we have to answer the question:\\

\textbf{Question: what is the underlying principle/mechanism that turns a TQFT into some combinatoric version of general relativity?}\\

This question has been partially investigated in Crane's paper~\cite{Crane:1995qj}. Based on the above reformulation of
Barrett-Crane model with the anyon-condensation mechanism and FRS formalism of RCFT,
I would like to suggest a possible answer to this. Actually a parallel
proposal for the microscopic mechanism of gauge theory has appeared long ago in the condensed matter community,
culminated in Wen's string-net condensation~\cite{Wen:2004}. Now I want to explain, similar picture applies to gravity more
straightforwardly.

We start with similar reasoning as in~\cite{Wen:2004}. As the subtitle indicates, we should first ask about the origin of the scalar
massless particles: phonons. We know for sure that phonons are not elemental particles, in contrast to photons. They emerge due to
the so-called spontaneously breaking of (continuous) symmetries. Then the Goldstone theorem tells us for each spontaneously breaking
symmetry, there is a massless Goldstone boson.  In the condensed matter community it seems more popular to use the phrase ``particle condensation''
to describe this mechanism. Namely instead of the original trivial vacuum, the system develops a new vacuum in which some of the orginal massive
excitations condense and break the corresponding symmetries. With respect to the new vacuum these excitations become massless. Since
continuous symmetries are described by Lie groups, we may say in a categorical way that particle condensation induces transition
between symmetric monoidal categories. Wen then go on to suggest that condensation of extended objects, the so-called string-nets,
could be the origin of light and fermions~\cite{Wen:2004}~\cite{Levin:2004mi}.

Although this is out of the scope of the present paper,
we want to say something about this proposal, which is commonly referred to as the Levin-Wen model of topological phases~\cite{Levin:2004mi}.
As the latter investigation shows, the full $(2+1)$D Levin-Wen model is equivalent to the Turaev-Viro theory
in the most general sense~\cite{Kirillov:2011}.
However, the generalization of Levin-Wen model from $(2+1)$D to $(3+1)$D is highly nontrivial. Obviously, it does not describe
the transition between symmetric monoidal categories, since that will lead to ordinary particle condensation. In my opinion,
$(3+1)$D Levin-Wen model realizes nontrivial vacuum of some unknown system described by certain braided monoidal $2$-categories, while using only the data of symmetric monoidal categories. This is a higher generalization of ordinary anyon condensation, in which one uses only the Frobenius algebras of a MTC to describe the $(2+1)$D nontrivial vacuum/condensed phase. The reason Levin and Wen could achieve this is that, they employ the Hamiltonian framework commonly used in the condensed matter community, which simplifies the data a lot. It is exactly this complicated vacuum that is supposed to generate the massless gauge bosons. Since our understanding of braided monoidal $2$-categories is
 still rather poor, the mathematical description of the original phase is not available.
This makes such a mechanism for the origin of light unacceptable at the moment. Only in the case when the involved categories
are finite, it is known to realize discrete gauge theories~\cite{Levin:2004mi}\cite{Walker-Wang:2011}.

Now let us turn to the discussion of gravity, and try to answer the above question in a similar way as for the gauge theory.
In a rather rough way, we may call a complete topological theory a solid phase, and, its nontrivial vacuum after condensation
a liquid phase. Then Wen's statement can be summarized as:\\

\indent\textbf{Wen's conjecture: the string-net liquids for Lie groups generate massless gauge theory.} \\

In a similar way, we conjecture:\\

\indent\textbf{Conjecture II: the string-net liquids for appropriate quantum groups generate gravity.}\\

 This is based on the following reasoning:\\

 \indent 1. As stated in \textbf{Conjecture I}, general relativity could be viewed as the result of a special diagonal anyon condensation.\\

 \indent 2. String-net liquids for quantum groups describe the nontrivial vacuum/condensed phase of diagonal anyon condensation~\cite{Kitaev:2011dxc}\cite{Kong:2013aya}.\\

\indent 3. The full $(2+1)$d Levin-Wen~(string-net) model is equivalent to
the Turaev-Viro theory in the most general sense~\cite{Kirillov:2011}.\\

\indent 4. Taking the classical limit of the Turaev-Viro theory, we reproduce the Ponzano-Regge formulation of 3D gravity. The stabilization from Turaev-Viro to Crane-Yetter theory is actually trivial, just as BF theory is a trivial bulk of the CSW theory.\\

\indent 5. Since both the $4$-cells and 3-cells are trivial after the condensation, the dynamical objects are the $2$-cells. Classically, these $2$-cells are described by the Regge calculus of general relativity.\\

In summary, the $2$-cells achieve dynamics after anyon condensation. When these dynamic surfaces move, split, and rejoin in the trivial ambient 4D space, gravity emerges. The string-net liquid is a dual, and vivid description of them.

\subsubsection{Cosmological Constant}

As a byproduct of the above reasoning, we may express the 4D cosmological constant through the lower dimensional parameters.
Following~\cite{Baez:1995ph}, it is not difficult to show that the 4D cosmological constant is related to the corresponding (boundary) CSW level $k$ as
\begin{equation}
k=\frac{12\pi}{\Lambda_4 l_4^2}=\frac{4\pi L_4^2}{l_4^2}.\label{eq:k4E}
\end{equation}
Clearly, now the coupling $k$ measures the cosmological area in the Planck units.
An interesting conclusion can be drawn if we combine the results for the 3D and 4D cosmological constant together. Notice that the same CSW theory appears both in the 3D and 4D gravity, as observed by Carlip in~\cite{Carlip:2015mea}. In particular, the relation (\ref{eq:k3E}) and (\ref{eq:k4E}) should be considered to be valid at the same time. Since $G_n$ and $l_n$ are by definition real and positive, requiring $\Lambda_3$, $k$, and $\Lambda_4$ be real
implies the following constraint:
\begin{equation}
\Lambda_3>0,\quad k>0,\quad \Lambda_4>0. \label{eq:CC3}
\end{equation}
Taking into account the quantization condition in~\cite{Witten:1988hf}, $k$ must actually be a positive integer.
This condition of $k$ is further reflected in the 2D theory, through the CSW-WZW correspondence~\cite{Witten:1988hf} or the FRS formulation mentioned above. Here the finiteness condition in the representation theory of $\widehat{\mathfrak{su}(2)}_k$ immediately constraints $k$ to be a positive integer~\cite{CFT}.
These results could be considered as the first indication that the 4D cosmological constant must be positive,
though here just for the Euclidean case.

\subsection{Lorentzian case}

The above results show that the 4D Euclidean gravity exhibits a very clear categorical structure, and can even be described quite
rigorously in the mathematical language. One may expect that similar results could be achieved in the Lorentzian case. However,
it turns out this expectation has never been met, even now. Remember that the 4D Euclidean gravity simplifies mainly due to the
accidental isomorphism (\ref{eq.Spin}). It is exactly this relation which allows us to express the 4D connection in terms of 3D ones,
and further relates the 4D gravity to the boundary CSW theory. Unfortunately, in the Lorentzian case such an accidental relation is lost. So how should we proceed without such an isomorphism?

\subsubsection{Identifying ``Time''}

In~\cite{Barrett:2000}, Barrett and Crane generalize their Euclidean model to the Lorentzian case. First they start with the 4D Lorentz group. Since no relation like (\ref{eq.Spin}) exists any more, they simply use instead its quantum deformation for the boundary theory. According to the general picture in~\cite{Baez:1995xq}, this is mathematically quite natural. However, the relation to the 3D symmetry is unclear. Later development attempts to project the 4D Lorentz group representations into 3D rotation group representations~\cite{Engle:2007}\cite{Freidel:2008},
which then encodes the 3D information. Here a very important question emerges:
should we take the boundary to be Euclidean, or Lorentzian instead?

If we follow the common quantization procedure for particles, it seems quite natural to take the boundary to be Euclidean. In the categorical language,
 the boundary space serves as the object, and the total spacetime the morphism. This is exactly the point made by Baez in~\cite{Baez:1999in}\cite{Baez:2004pa},
 in parallel with the state/process in quantum mechanics. However, as we discussed in the previous section, gravity exhibits some novel features
 not appearing in the particle picture. In short, it should better be treated in an extended framework, rather than an ordinary category.
 In particular, the reduction from 4D to 3D is topological and trivial. As a manifestation of this, when we reduce the 4D theory
 to 3D we get an Hamiltonian constraint, instead of a dynamical Hamiltonian. Thus we get the answer, the dynamics of gravity should be preserved in the boundary, which should be taken to be Lorentzian.

 Therefore we should abandon the approaches in~\cite{Engle:2007}\cite{Freidel:2008}, and all the nice features of the Hilbert space in the
  canonical formulation. Moreover, since (\ref{eq.Spin}) is an isomorphism, we could start by Wick rotating either the left-handed side, or the right
-handed side. Obviously Ashtekar and Barrett-Crane have chosen to Wick rotate the left-handed side, and start from the 4D bulk theory.
 But from our discussion in the previous section, we see that the 3D theory is more important and plays the key role. Therefore we are
 finally led to the choice of the symmetry group $\text{SO}(2,1)\times \text{SO}(2,1)$, or its quantum deformed version.
 The latter may not be well defined mathematically yet. However, thinking of the equivalence (\ref{eq:3-2}), one could also start from
 the affine Lie algebra $\widehat{{\mathfrak s \mathfrak o}(2,1)}_k\otimes \widehat{{\mathfrak s \mathfrak o}(2,1)}_k$, which is
 well-defined in the Lorentzian case, at least physically. When the symmetry group is settled done as this, all the formulation
 in the Euclidean case can be imitated. First we build a Lorentzian doubled CS theory in 3D. To further include 2d boundaries/defects,
 we select a Frobenius algebra as before
  \begin{equation}
 A=R_\alpha \otimes  R_\alpha^\vee, \label {eq:A3}
 \end{equation}
which is considered as an object in the representation category of $\widehat{{\mathfrak s \mathfrak o}(2,1)}_k$.
Here the difference from the previous $\widehat{{\mathfrak s \mathfrak u}(2)}_k$ appears. As we show before,
for $\widehat{{\mathfrak s \mathfrak u}(2)}_k$
the different choices of ``A'' are limited to a finite set.
But for $\widehat{{\mathfrak s \mathfrak o}(2,1)}_k$, it seems that we may
have a continuous spectrum. This is indeed the case if we take a Lie algebra point of view,
similar as in Barrett-Crane's analyses~\cite{Barrett:2000}.
However, from a VOA point of view, the spectrum is still discrete. The underlying reason is the Virasoro conditions.
A famous example is the discrete string mass spectrum~\cite{GSW:1987}. I made this comparison in the recent note~\cite{Zuo:2016ezr} in the limit $k\to \infty$,
following Carlip~\cite{Carlip:1994gy}. The spectrum may even be finite when $k$ is finite, similar as the $\widehat{{\mathfrak s \mathfrak u}(2)}_k$
case. The 3D-2d relation is again established by deriving the
corresponding center, or diagonal modular invariant:
\begin{equation}
Z(A)=\oplus_\alpha R_\alpha\times R_\alpha^\vee, \label{eq:ZA2}
\end{equation}
With $\alpha$ taking discrete values, this should be well-defined. Again when restrict the 3D theory to the vacuum $Z(A)$, and the 2d boundaries/defects to $A$'s,
we have anyon-condensed the corresponding 4D BF theory to general relativity.

\subsubsection{Emergence of 4D Lorentz Symmetry}

One may doubt this procedure by saying: but you have never started from the 4D theory at all! Yes, we have done that. When we
formulate the Lorentzian doubled CS theory, we are secretly dealing with the corresponding BF theory. The real question is:
how is the physical 4D Lorentz symmetry, rather than the 3D symmetry we have chosen, reflected?

At this point we should distinguish two points: a local symmetry and gauging of a local symmetry. What is really done in Ashtekar's
original formulation is to gauging the 4D Lorentz symmetry. General relativity is recovered when we impose Lorentz invariance condition
at the last step. This procedure is then followed in the later studies, e.g. in Barrett-Crane's Lorentzian spin networks. What we are proposing
here is that, since the 4D-3D relation is topological, what we really need to do is gauging the 3D symmetries. The only condition we has to
impose is that, the 4D Lorentz invariance be recovered finally. This is obvious: although the representation category of $\widehat{{\mathfrak s \mathfrak o}(2,1)}_k$
may not be modular, after anyon condensation the 3D theory has a unique state, the vacuum (\ref{eq:ZA2}). Therefore its stabilization into 4D gives a trivial center, which is simply
the Lorentz singlet. This is succinctly stated as: ``the bulk of a bulk is trivial''~\cite{Kong:2014qka}. In other words,
the 4D Lorentzian symmetry is emergent. A direct corollary of this discussion is that, Carlip and Strominger's derivation of the 3D black hole entropy~\cite{Carlip:1994gy}~\cite{Strominger:1997eq}, actually apply secretly to the physical 4D black holes. This is the underlying proposal made in may recent note~\cite{Zuo:2016ezr}. Unfortunately,
the issues with the local symmetries and the gauging choice is never spelled out there.

In summary, although we don't have an analogous isomorphism as (\ref{eq.Spin}) in the Lorentzian case, we have the trivial homomorphism from
(\ref{eq:ZA2}) to the 4D Lorentz singlet, which is just enough for our construction of physical gravity, at least at the 4D-3D level. Similar idea was
proposed in~\cite{Barenz:2016nzn}. Later we will show that when the full framework is established, an exact isomorphism for the respective categories
as (\ref{eq.Spin}) may indeed exist, in the form of some extended bulk-boundary duality.

\subsubsection{Cosmological Constant}

With all these issues clarified, we could now employ this framework to discuss the question related to the physical cosmological constant
$\Lambda_4$. Actually, since all the elements are similar to the Euclidean case, the derivation of $\Lambda_4$ could be done in exactly the same way, as in~\cite{Baez:1995ph}. We immediately get the same relation (\ref{eq:k4E})
 \begin{equation}
k=\frac{12\pi}{\Lambda_4 l_4^2}=\frac{4\pi L_4^2}{l_4^2}.\label{eq:k4L}
\end{equation}
The relation does not change simply because the fourth dimension is spatial, so the 4D-3D relation is not affected by the Wick rotation
in the remaining 3 dimensions. We recall here that the coupling $k$ is further related to the 3D cosmological constant as (\ref{eq:k3L})
\begin{equation}
k=\frac{1}{4G_3\sqrt{-\Lambda_3}}.
\end{equation}
Now we impose the reality condition on $\Lambda_3$, $k$, and $\Lambda_4$, and obtains
\begin{equation}
\Lambda_3<0,\quad k>0,\quad \Lambda_4>0.
\end{equation}
Comparing to (\ref{eq:CC3}) we see that while $\Lambda_3$ changes its sign after the Wick rotation, $\Lambda_4$ remans positive. Therefore we end up with the
simple conclusion:\\

\indent \textbf{Surrounded by a 3 dimensional Anti-de Sitter spacetime, is always a 4 dimensional spacetime with a positive
cosmological constant, i.e., a de Sitter spacetime in general. }\\

Be careful that, since the 3D gravity is required to be ``doubled'',
the bounding is actually an embedding. Note that similar derivation has been done recently in the series of papers~\cite{Haggard:2014xoa}\cite{Haggard:2015yda},
but with a real Barbero-Immirzi parameter $\gamma$~\cite{Barbero:1994ap}\cite{Immirzi:1996dr}. If one insists on the original Ashtekar variables (with $\gamma=\pm\mathi$), the above result, originally due to Baez~\cite{Baez:1995ph}, is recovered.

\section{Categorification}
With all the previous discussion we may say, we have fully explained Crane's early conjecture:\\
\indent Crane's conjecture I: ``The universe is in the CSW state''~\cite{Crane:1993vs}.\\
Now we turn to Crane's further conjecture:\\
\indent Crane's conjecture II: ``A categorification of the CSW theory''
provide ``a quantum theory of gravity''~\cite{Crane:1995qj}.

\subsection{Motivation}
First we want to review Crane's motivation for such a suggestion. The basic reason is that, as we mentioned before,
the reduction from 4D to 3D is a topological one, since the Hamiltonian degenerates to a constraint~\cite{Smolin:1995vq}.
Therefore, at the level of 4D-3D, we lost the dynamics and the resulting 3D CSW theory is thermal~\cite{Smolin:1994qb}\cite{Connes:1994hv}.
It is expected that by categrifying CSW theory, the dynamics could be regained~\cite{Crane:1995qj}. Moreover, decategorifying the
resulting theory naturally leads to a thermal state.

I must say the reason is not so convincing, at least for me. We have argued that even if the 4D-3D reduction is topological,
the dynamics could be restored in the remaining spacetime. I will use this opportunity to give a categorical view on space and
time. In~\cite{Baez:1999in}\cite{Baez:2004pa}, Baez has proposed that spacetimes naturally forms a category, with spaces the objects
and spacetimes the morphisms. In other words, the morphous structure could be viewed as a (pre)-temporal structure. But how about the
spatial directions? A natural answer is that monoidal structure gives rises to the spatial structure. With this general picture in mind,
we may say Baez and Dolan's $k$-tuply monoidal $n$-cateories~\cite{Baez:1995xq} $n\mbox{Cat}_k$ describe a general exotic spacetime with $k$ spatial dimensions and $n$
temporal dimensions~\footnote{We use the same notation as~\cite{Baez:1995xq} and keep the parameter $k$ here. Be careful not
to confuse it with the CSW coupling $k$.}. Therefore, the 4D-3D reduction from BF theory to CSW theory indeed reduces spatial dimension by one,
since from the categorical view
it corresponds to a deformation from symmetric monoidal categories to braided monoidal categories. However,
in the resulting spacetime we still have a time variable, just because we still have morphous structure.
So we must find some other physical motivation for catgorification, if we believe Crane's proposal is the correct one.

But there are so many reasons that we must do this! Here we list three of them, which are closely related to the content of the paper.

First, we can immediately give a similar but much stronger reason than Crane's: we have black holes in general relativity.
In addition to gravity dynamics, black holes are also thermal and have huge entropy. Based on the same reasoning
as in~\cite{Crane:1995qj}, we would expect that by categorifying the present formulation, the microscopic states of black holes could
be reproduced. Actually, most successful interpretations of black hole entropy, such as the famous approach in~\cite{Strominger:1996sh},
and the two derivations mentioned previously~\cite{Carlip:1994gy}\cite{Strominger:1997eq},
inevitably use some kind of 2d CFT. Even the more recent attempt to describe directly Kerr black holes
uses some structure from 2d CFTs~\cite{Guica:2008mu}. But as we know, even the simplest RCFTs can not
be accommodated by the $1\mbox{Cat}_k$ column in the periodic table of~\cite{Baez:1995xq}: according to the FRS framework~\cite{Fuchs:2001am}\cite{Runkel:2005qw}, MTCs only capture the global structure
of RCFTs. To fully describe them, we need suitable VOAs as the local data.
This bifurcating description is certainly not so satisfying. A natural guess would be that,
the local data and global structure of 2d CFTs, rational or not, could be treated in a unified framework
in the $2\mbox{Cat}_k$ column~\cite{Baez:1995xq}. We thus see that both the physical and mathematical developments demand categorification.

The second reason originates from a careful inspection on the cosmological constant and the Planck constant
from the categorical point of view. As we mentioned, this is the primordial motivation of this paper.
According to~\cite{Baez:1995xq}, we can deformation quantize the symmetric monoidal categories and get braided monoidal categories. The deformation parameter $q$ should somehow be determined by the real CSW coupling $k$. So this deformation quantization procedure involves a single real parameter $k$. As we have shown previously, $k$ is further related to the cosmological constant. Due to this, Baez suggested to call quantum groups
 ``cosmological groups''~\cite{TWF:183}.  But as a quantization procedure, it should also be related to the Planck constant $\hbar$. The truth is, since both $q$ and $k$  are dimensionless, they must be given by some dimensionless
 combination of the Planck length and the cosmological constant~\cite{TWF:183}. In other words, we have a single real parameter $k$, which is not enough to provide the origin of both the Planck constant and the cosmological constant. Only some special combination of the latter two is reflected in the deformation parameter. In order to produce the Planck constant and the cosmological
 constant independently, we need a deformation procedure involving two real parameters. A natural guess is the corresponding deformation
 quantization in the column $2\mbox{Cat}_k$~\cite{Baez:1995xq}. Therefore, we need categorification.

The third reason is more technical. We have shown that the crucial step in bringing topological BF theory to
general relativity is to anyon condense diagonally the doubled MTC to the Frobenius algebra, or diagonal
modular invariant. The key step in this anyon condensation process is to find a Frobenius object in a
monoidal category~\cite{Kong:2013aya}. A systematic way to achieve this is to promote the monoidal category to a $2$-category, and then perform the ``adjunction'' operation~\cite{TWF:89}\cite{TWF:174}. Our special Frobenius algebra (\ref{eq:A}) could be reproduced easily in this way~\cite{Schweigert:2006af}\cite{Fuchs:2012dt}. Historically, a ``classical'' limit of such a procedure has led to the birth of string theory. Indeed, it is the conjectural string scattering amplitudes which naturally satisfy the crossing symmetry required by the dual model of hadrons. By a similar reasoning as string theory, we would call for categorification now.

\subsection{General framework}

Now we try to establish the framwork for categorifying CSW theory as Crane proposes. Essentially, this means to categorify
braided monoidal categories, and most importantly, quantum groups. In~\cite{Crane:1994ty}, the authors even sketched
a rough framework for such a categorification. Their proposal could be viewed as a higher version of the Tannaka-Krein
 reconstruction theorems~\cite{JS:1990}. The original reconstruction theorems give a duality between certain Hopf algebra and the braided
 monoidal category of its representations. Crane and Frenkel believe that a categorification of this duality would
 involve a triality: certain ``trialgebra'', a kind of ``Hopf category'' of its representations, and some kind of
 monoidal bicategory of its 2-representations.

\subsubsection{Vertex Operator Algebras}

Here I want to make some comments on these concepts. We now know from Baez and Dolan's periodic table that
categorification of braided monoidal categories ($1\mbox{Cat}_2$) should be some kind of braided monoidal $2$-categories~
($2\mbox{Cat}_2$). But it is quite unclear, at least for me, how certain trialgebras or Hopf categories would categorify
bialgebras and Hopf algebras. Some physical picture would be helpful here. We know all the ordinary algebra structure, such
as algebras/coalgebras, bialgebras/Hopf algebras, could be defined using string diagrams. Physically these diagrams could be viewed as worldlines of particles.
A categorification of these worldlines should be some worldsheets of some hypothetical strings. This means, the expected higher algebra structure
should be defined on two-dimensional surfaces, typically Riemann surfaces. Therefore one should expect the higher version
of bialgebras/Hopf algebras be some kind of VOAs, or variants thereof. The Kazhdan-Lusztig equivalence (\ref{eq:3-2}) gives some indirect
evidence for this hypothesis. In particular, one should not expect two separate levels of some kind of Tannaka-Krein duality,
but a single higher version of Tannaka-Krein duality.

\subsubsection{Theory X and 4D topological invariants}

Additional doubts on Crane and Frenkel's proposal come from its poor relation to the novel topological invariants of 4D manifolds.
We know that in the $1\mbox{Cat}_k$ column, quantum groups, or braided monoidal categories of their representations,
immediately give us novel topological invariants for knots/3-manifolds~\cite{Reshetikhin:1991}. In the $2\mbox{Cat}_k$ column, the
situation is still quite unclear. A short review of the status and progresses achieved so far is given in~\cite{Gukov:2017}.
Here we will give a slightly different review on this, partly based on Moore's lecture notes~\cite{Moore:2012} and the notes
from~\cite{Ben-Zvi:2014}.

We start from the stabilization case, namely the symmetric monoidal $2$-categories ($2\mbox{Cat}_4$). In the $1\mbox{Cat}_k$ column, we know the 3D CSW theory could be considered as the boundary theory of a 4D BF theory with a cosmological constant.
Formally we could take the classical limit by taking the cosmological constant to zero. The resulting pure BF theory would depend only on the ``classical'' Lie algebra, or the symmetric modoidal category of its representation. One would expect a similar classical theory in the $2\mbox{Cat}_k$ column, related to a Lie $2$-algebra, or the symmetric monoidal $2$-category of its $2$-representations. In~\cite{Baez:2003fs} (see also the talk~\cite{Baez:2007}) the basic structure of a Lie $2$-algebra is analyzed, and shown to reduce to a Lie algebra together with some extra structure parameterized by a real number $k$. This makes it very natural to relate a Lie $2$-algebra to an affine Lie algebra at level $k$, when $k$ is an integer. However, no interesting 6d theory has been constructed from the Lie $2$-algebras.

It turns out that the most plausible candidate is the mysterious 6d (2,0) superconformal theory~\cite{Witten:1995zh}
\cite{Strominger:1995ac}. More precisely, what we really want is certain holomorphic-topological twist of the (2,0) theory,
which is called ``Theory X'' by some mathematicians~\cite{Ben-Zvi:2014}. Theory X depends only on a simple Lie algebra, most
naturally of ADE type. Probably we could consider the underlying algebra structure as some classical limit of
Baez and Crans's Lie $2$-algebra~\cite{Baez:2003fs}. Therefore it is completely parameter free.

Many efforts have been made on the (2,0) theory/Theory X, but it still remains quite a mystery. Some properties of it are listed
in~\cite{Moore:2012} as axioms. However, if we consider it as a higher analogue
of the classical BF theory, then the theory itself should not be quite interesting. What we should be concerned with is the
theory after some $2$-deformation quantization, in the sense of Baez and Dolan~\cite{Baez:1995xq}. While the deformation quantization of
classical BF theory gives CSW theory in 3D, here we would expect some kind of $2$-deformation quantization of the
6d (2,0) theory. The resulting theory should live in 4D, and thus give us novel topological invariants of 4-manifolds.

This is indeed the case. Physically, we dimensionally reduce the 6d theory on a punctured Riemann surface
$\mathcal{C}$ with some special twist~\cite{Witten:1988ze}. While the original twist proposed in~\cite{Witten:1988ze} is completely topological,
here the special twist makes the theory depend holomorphically on the Riemann surface, and thus called ``holomorphic twist''.
The procedure results in a special kind of $\mathcal{N}=2$ supersymmetric gauge theories in 4D, called ``class S''~\cite{Gaiotto:2009we}. One can further apply various
topological twists to these class S theories, and obtain interesting 4D TQFTs and topological invariants. We want to mention that, the holomorphic twist and all the topological twists can be considered as specifying some
special bundle structure on the corresponding manifolds, as explained in~\cite{Gadde:2013sca} and recently reviewed in~\cite{Dabholkar:2020fde}. Depending on the topology of the Riemann surface and the 4D topological twists applied, we have:

1. When $\mathcal{C}=\mathbb{S}^2$ with punctures~\cite{Gaiotto:2009we}, we can apply the same topological twists as in~\cite{Witten:1988ze}, and obtain
Donaldson-Witten like topological invariants~\cite{Donaldson:1983An}\cite{Witten:1988ze}.

2. When $\mathcal{C}=\mathbb{T}^2$, the supersymmetry is enhanced to $\mathcal{N}=4$. There are three different topological twists
we can employ:  \\
\indent a) the Donaldson-Witten twist~\cite{Witten:1988ze};\\
\indent b) the geometric-Langlands twist~\cite{Marcus:1995mq}\cite{Kapustin:2006pk}, which gives Kapustin-Witten topological
theory related to the geometric Langlands dualtiy~\cite{Kapustin:2006pk};\\
\indent c) the Vafa-Witten twist~\cite{Vafa:1994tf}, which gives the Vafa-Witten theory and the corresponding Vafa-Witten invariants~\cite{Vafa:1994tf}.

3. When $\mathcal{C}=\mathbb{D}^2$, one may also apply the Vafa-Witten twist, and get the Gaiotto-Witten topological theory~\cite{Witten:2011zz}
\cite{Gaiotto:2011nm} encoding Khovanov homology~\cite{Khovanov:1999qla}.

So we indeed get lots of novel invariants for 4-manifolds. But how do we know that the whole framework indeed
reflects the structure of $2\mbox{Cat}_k$~\cite{Baez:1995xq}? Though not completely sure, we have
two quite compelling evidences for this: \\
\indent First, the reproduction of Khovanov homology from Gaiotto-Witten construction~\cite{Witten:2011zz}
\cite{Gaiotto:2011nm}. Khovanov homology is a categorification of the Jones polynomial for knots, which could be
alternatively derived from the CSW theory~\cite{Witten:1988hf}. Therefore, the Gaiotto-Witten theory properly
encodes Khovanov homology strongly suggests that we are on the right way of categorifying CSW theory.

Secondly, the reduction of Theory X to 4D is naturally accompanied by a two-parameter deformation called
$\Omega$-deformation~\cite{Nekrasov:2002qd}\cite{Nekrasov:2003rj}\cite{Nekrasov:2009rc}. Although a rigorous mathematical defintion of the $\Omega$-deformation
 has never been given till now, there are plenty of results~\cite{Yagi:2014toa}\cite{Beem:2018fng}\cite{Oh:2019bgz}\cite{Jeong:2019pzg}
indicating that it could be viewed as $2$-deformation quantization for symmetric monoidal $2$-categories ($2\mbox{Cat}_4$)~\cite{Baez:1995xq}.
A similar formulation of this identification using $E_n$-algebras, rather that $n$-categories, is recently demonstrated in~\cite{Beem:2018fng}.
In the following we will simply assume this identification and take the two deformation parameters as $\epsilon_1$ and $\epsilon_2$.

\subsubsection{VOA[$M_4$] and $2$-Tannaka-Krein duality}

Although we got lots of 4D topological invariants from the reduction of Theory X, we are still far from solving our
categorifacation problem. This is because the formulation of these invariants heavily relies on supersymmetric quantum field theories,
which are very hard to define rigorously. We want a combinatorial framework based directly on the structure of the braided
monoidal $2$-categories ($2\mbox{Cat}_2$)~\cite{Baez:1995xq}, just like the Reshetikhin-Turaev formulation of CSW theory.
This seems to be extremely difficult to achieve directly.
Crane and Frenkel proposed that these $2$-categories could manifest themselves as $2$-representations of trialgebras or Hopf
categories. We have explained that, from the physical picture, maybe VOAs are the suitable algebra structure
to start with. So we may expect some higher Tannaka-Krein duality to show up here, which would give us nice $2\mbox{Cat}_2$'s
from the $2$-representations of VOAs. Still this is rather difficult. But if this is indeed true, we should also be able to
get those 4D invariants directly from the corresponding VOAs. Thus, we should seek the relation between VOAs and the (twisted)
supersymmetric field theories of class S, and further the relation between VOAs and 4-manifolds. It turns out that there are
a lot of such relations, in contrast to the situation of trialgebras and Hopf categories.

The existence of some relation between 2D field theories and 4D field theories is anticipated at the end of~\cite{Atiyah:1987ri}. In the last 20 years various explicit relations for concrete theories have been found. These include the BPS/CFT correspondence
proposed around 2003~\cite{Losev:2003py}~(reviewed in~\cite{Nekrasov:2015wsu}), the Alday-Gaiotto-Tachikawa~(AGT) correspondence found in 2009~\cite{Alday:2009aq},
and the 4D Supersymmetric Conformal Field Theory~(SCFT)/VOA duality deduced in 2013~\cite{Beem:2013sza}\cite{Beem:2014rza}. However, all of them involve the 4D supersymmetric quantum
field theories, thus hard to define rigorously. Since these theories could produce 4-manifold invariants after topological twisting,
it is natural to eliminate these field theories and build the duality directly on 4-manifolds. Also the structure of 2d CFTs,
if not rational, are not understood quite well. One would like to restrict to its local data, the VOAs, which could be defined
rigorously. Linking these two parts together constitutes the motivation of the series
of papers by Gukov and his collaborators~\cite{Gadde:2013sca}\cite{Gukov:2017}\cite{Gukov:2018}.

The main result of these papers is the duality between certain VOAs and 4-manifolds. Explicitly, the duality reads:
\begin{equation}
\mbox{VOA} \quad \Longleftrightarrow \quad [M_4, \mathfrak{g}]_{\epsilon_1,\epsilon_2}.\label{2TK}
\end{equation}
The simple Lie algebra $\mathfrak{g}$ of ADE type originates from the 6d (2,0) theory. The parameters $\epsilon_1$ and $\epsilon_2$
are due to the $2$-deformation quantization. As for the information one must input for the $4$-manifold $M_4$, it is not quite clear.
My personal understanding is that, one only needs to specify the classical topological information, like the homology/homotopy data,
for $M_4$. In the language of~\cite{Baez:1998}, one needs only the groupal data. Again here the comparison with the $1\mbox{Cat}_k$ column
is very helpful. In that case, we could color the classical $3$-manifold $M_3$ (with or without knots/links) with the quantum group $U_q\mathfrak{g}$, and
make it quantum. In a similar way, here the right-handed side of (\ref{2TK}), $[M_4, \mathfrak{g}]_{\epsilon_1,\epsilon_2}$, denotes a $2$-quantum $4$-manifold, with the full monoidal data~\cite{Baez:1998}. Thus it is a manifestation of the underlying braided monoidal $2$-category ($2\mbox{Cat}_2$),
which itself is difficult to construct at the moment.
Interpreted in this way, the duality found by Gukov et al. could be considered as a ``$2$-Tannaka-Krein duality''.

Let us take one simple example from~\cite{Gukov:2018} to illustrate all these statements. When $\mathfrak{g}=\mathfrak{sl_2}$ and
$M_4=\mathbb{R}^4$, we have~\cite{Alday:2009aq}:
\begin{equation}
\mbox{VOA}[\mathbb{R}^4, \mathfrak{sl_2}]_{\epsilon_1,\epsilon_2}=\mbox{Vir}_b,\label{eq:Vir}
\end{equation}
with $\mbox{Vir}_b$ the Virasoro algebra for the Liouville quantum field theory. The explicit relation between the parameters on
the two sides are~\cite{Alday:2009aq}\cite{Teschner:2010je}\cite{Gukov:2018}:
\begin{equation}
\epsilon_1=\hbar b,\quad \epsilon_2=\hbar/b.\label{eq:b}
\end{equation}
Here $b$ is the usual parameter in Liouville theory, and the Planck constant $\hbar$ denotes the ordinary quantization of the theory.
This simple example is then used as a patch to build more general $4$-manifolds, and the corresponding VOAs~\cite{Gukov:2018}.

\subsubsection{Gravity and Cosmological Constant}

Thus we have obtained the main framework of categorifying the $1\mbox{Cat}_k$ structure to the $2\mbox{Cat}_k$ structure.
As Crane proposed~\cite{Crane:1995qj}, this should be related to a quantum theory of gravity. But how is gravity reproduced?
Formally, this would involve the inverse procedure of ``de-categorification''~\cite{Crane:1995qj}\cite{Baez:1998}. Physically de-categorification would mean some kind of ``trace'' over a circle, which would naturally induce gravity thermodynamics~\cite{Crane:1995qj}. To actually perform this de-categorification operation, we would like to relate both the $1\mbox{Cat}_k$ and $2\mbox{Cat}_k$ to some VOAs, or the corresponding CFTs.

In the last section, we have shown that general relativity could be represented through a diagonal anyon condensation. As a
result, Euclidean gravity is determined by a special Frobenius algebra in a proper MTC. Similar structure is expected in the Lorentzian case, with the MTC replaced by the representation category of $\widehat{{\mathfrak s \mathfrak o}(2,1)}_k\otimes \widehat{{\mathfrak s \mathfrak o}(2,1)}_k$.
This is the VOA controlling the $\text{SO}(2,1)$ WZW CFT. From the FRS framework~\cite{Fuchs:2001am}, we may say the global structure of this WZW theory determines general relativity.

From the discussion in this section, we know that the information of the $2\mbox{Cat}_k$ structure is encoded also in some VOAs, with the basic element the Virasoro algebra $\mbox{Vir}_b$, which encodes the Liouville CFT. To link the two parts together, we must find out the relation between these two field theories. As we mentioned before, these two are actually equivalent to each other, at the level of correlation functions~\cite{Teschner:2005}\cite{Schomerus:2007}\cite{Giribet:2008}.
Here I will not repeat the derivation or proofs in these papers. Instead, I will give an intuitive explanation why these two
would be equivalent.

The argument goes like this. Before anyon condensation, the 3D CSW theory, or 3D gravity is completely topological. The
only solution, if not allowed to have horizons (finite boundaries), is pure $\text{AdS}_3$ spacetime. Such a spacetime possesses $\text{SO}(2,1)\times \text{SO}(2,1)$ isometry. Now consider the physics at the asymptotic boundary. There are two different ways to study it. We may first gauging the isometry, and get the bulk CSW description. Then we restrict the CSW theory to the asymptotic boundary, which gives rise to the $\text{SO}(2,1)$ WZW theory~\cite{Witten:1988hf}\cite{Carlip:1994gy}. Alternatively,
we notice that the isometry of $\text{AdS}_3$ induces a global conformal symmetry $\text{SO}(2,2)$ on the boundary. Properly ``gauging'' this symmetry on the 2D boundary, we obtain the Liouville field theory~\cite{Brown:1986nw}\cite{Coussaert:1995zp}\cite{Strominger:1997eq}. Since the bulk theory is topological, these two procedures should be equivalent. Therefore the two CFTs should be equivalent, which may be called Liouville-WZW
correspondence~\cite{Teschner:2005}\cite{Schomerus:2007}\cite{Giribet:2008}. A manifestation of this equivalence
is that both of them produces correctly the Bekenstein-Hawking formula for the entropy of 3D black holes~\cite{Carlip:1994gy}\cite{Strominger:1997eq}.

There is even a third picture of the $\text{AdS}_3$ spacetime.
We know that $\text{AdS}_3$ is naturally defined in an exotic 4D spacetime with an extra ``time'', with its isometry originating from
the indefinite ``rotational'' symmetry $\text{SO}(2,2)$ of this exotic 4D spacetime. Remembering that ``time'' is always interpreted as ``morphous'' direction in the categorical
picture, we see that $\text{AdS}_3$ is secretly a manifestation of a $2\mbox{Cat}_2$ structure. Therefore it also serves as a bridge linking two sides of the $2$-Tannaka-Krein duality~(\ref{2TK}), similar as the reasoning at the end of~\cite{Atiyah:1987ri}. Moreover, here we finally see the Lorentzian analogue of the isomorphism (\ref{eq.Spin}):
\begin{equation}
\text{SO}(2,2) \cong \text{SO}(2,1)\times \text{SO}(2,1),\label{eq.SO22}
\end{equation}
up to spin factors. Due to this isomorphism, our previous choice of the gauge group $\text{SO}(2,1)\times \text{SO}(2,1)$ naturally emerges from Crane's categorification program. Note that this isomorphism has already be used in~\cite{Witten:1988} to study 3D gravity.

Now we have all the information to reproduce realistic gravity from the $2$-categorical structure.
It would be quite helpful to employ the ``collapsing'' procedure proposed by Soibelman~\cite{Soibelman:1900tm}. As illustrated in~\cite{Kong:2011jf}, 2d CFTs could be considered as a ``$2$-spectral geometry''. If we collapse this
$2$-spectral geometry to a graph field theory, we should get gravity~\cite{Soibelman:1900tm}. Here we could make this collapsing picture much more explicit. First we use the $2$-Tannaka-Krein duality to reduce the $2\mbox{Cat}_k$'s to the corresponding VOAs, with $\mbox{Vir}_b$ the elementary data. Correspondingly, we could start with the Liouville filed theory. Then this could be related to the $\text{SO}(2,1)$ WZW theory, through the Liouville-WZW correspondence. Finally we hide the local data of this WZW theory, keeping only its global structure. This would give rise to
physical gravity with the cosmological constant, according to~\cite{Barrett:1997gw}\cite{Zuo:2017hii} and the discussion in the previous section. The hidden local data, namely the $\widehat{{\mathfrak s \mathfrak o}(2,1)}_k\otimes \widehat{{\mathfrak s \mathfrak o}(2,1)}_k$ affine Lie algebra, manifests itself as the hidden degrees of freedom for the black holes~\cite{Carlip:1994gy}.

The key relation in the Liouville-WZW correspondence is (\ref{eq:kb}), which we repeat here
\begin{equation}
k-2=b^{-2}.\label{eq:kb1}
\end{equation}
Combining (\ref{eq:kb1}), (\ref{eq:b}) and (\ref{eq:k4L}), we could completely determine the Planck constant $\hbar$ (or Planck length $l_4$) and the cosmological constant $\Lambda_4$ from the $2$-deformation quantization parameters $\epsilon_1$ and $\epsilon_2$. This is a strong evidence that the whole $2$-categorical structure naturally forms the foundation of quantum gravity.

\section{Discussion}

In this final section we would to list some conceptional issues related to the contents of this paper: \\

1. As for black holes: \\

While various black holes may correspond to quite different VOAs, probably all of them could be produced
by gluing pieces of Virasoro algebras $\mbox{Vir}_b = \mbox{VOA}[\mathbb{R}^4_{\epsilon_1,\epsilon_2}]$~\cite{Gukov:2018}.
Therefore, the microscopic states could be universally accounted for either as in~\cite{Carlip:1994gy}, or equivalently as in~\cite{Strominger:1997eq}.\\

2. As for the cosmological constant:\\

 The cosmological constant and the Planck constant are completely independent of each other in the present framework.
Hence, the vacuum energy of quantum particles has nothing to do with the cosmological constant. So the naive estimation
of the cosmological constant from the vacuum energy of known particles is ridiculous.\\

3. As for the loop formulation of gravity:\\

Categorification of the current loop formulation of gravity is extremely urgent. In particular, replacing the local Hopf-algebra data by proper VOAs would be the key step.\\

4. As for string theory:\\

\indent a) String theory has struggled for a long time to provide a natural explanation of the realistic cosmological constant. Compared to the derivation in the present paper, it becomes very clear that this is because current formulation of string theory always works in the stabilization limit. This is true for the critical string theory, or its reduction to the 6d (2,0) theory. To accommodate the cosmological constant, the deformation to lower dimensions is inevitable.\\

\indent b) The common conclusion in the string theory community is that gravity dynamics is completely decoupled in the 6d (2,0) theory. Therefore the theory has nothing to do with quantum gravity. In this paper we have illustrated how some twisted deformation of this theory perhaps encodes the whole information of quantum gravity. If this turns out to be true, then the standard physical interpretation of current string theory framework should be revised.\\

\section*{Note added:}

When I was finishing this paper, I learned accidentally from the conference note \cite{Frenkel:2012} that Igor Frenkel
had many contributions to representation theory closely related to the contents here. These include, in addition to Vertex Operator Algebras and
Categorification, his ideas and work on double loop groups, on split real quantum groups, and many others. I hope that physicists
would, jointly with mathematicians, study Frenkel's ideas carefully, and further develop those frameworks he proposed.
I believe this would help quite a lot establish the quantum theory of gravity, through the procedure sketched in this paper.

\section*{Acknowledgments}
I would like to thank Si Li for encouraging me to study Baez and Dolan's pioneer paper ``Higher-Dimensional Algebra and
Topological Quantum Field Theory''~\cite{Baez:1995xq} a few years ago, and Yi-hong Gao for suggesting me to revisit the Alday-Gaiotto-Tachikawa
correspondence from the categorical viewpoint. I am particularly grateful to Liang Kong for supporting me to attend his course
``Tensor categories and 2d topological orders'' during the period 2017.9.19--2017.10.19. Some ideas here originated
from the discussions with him at that time.

\end{document}